\documentstyle[twocolumn,prl,epsf,aps]{revtex}

\newcommand{\rmd}{{\rm d}}
\newcommand{\rmi}{{\rm i}}

\newcommand{\beq}{\begin{equation}}
\newcommand{\eeq}{\end{equation}}
\newcommand{\bea}{\begin{eqnarray}}
\newcommand{\eea}{\end{eqnarray}}

\newcommand{\cuoo}{CuO$_{2}$}
\newcommand{\oco}{O-Cu-O}
\newcommand{\lco}{La$_{2-x}$Sr$_{x}$CuO$_{4}$}

\newcommand{\ybco}{YBa$_{2}$Cu$_{3}$O$_{6+x}$}

\begin{document}                                                

\wideabs{

\draft

\title{Quantum skyrmions and the destruction
of long-range antiferromagnetic order in the high-T$_c$
superconductors {\lco} and {\ybco}}

\author{Eduardo C. Marino and Marcello B. Silva Neto}

\bigskip

\address{
Instituto de F\protect\'\i sica, Universidade Federal do Rio de Janeiro,
Caixa Postal 68528, Rio de Janeiro - RJ, 21945-970, Brazil
}

\date{\today}
\maketitle


\begin{abstract} 

We study the destruction of long range antiferromagnetic 
order in the high-T$_c$ superconductors {\lco} and {\ybco}. 
The CP$^{1}$-nonlinear sigma model formulation of the 
two-dimensional quantum Heisenberg antiferromagnet is used for describing
the pure system. Dopants
are introduced as independent fermions with an appropriate
dispersion relation determined by the shape of the Fermi surface. 
Skyrmion topological defects are shown to be
introduced by doping and their energy is used as an order parameter for the 
antiferromagnetic order. We obtain analytic expressions for the
skyrmion energy as a function of doping which allow us to plot,
without adjustable parameters,
the curves $T_{N}(x_{c})\times x_{c}$ and $M(x)\times x$, for 
the two compounds, in good quantitative agreement with the 
experimental data.

\end{abstract}


} 

\begin{narrowtext}


High-temperature superconductivity is by now well established 
to arise from doping quasi two-dimensional (quasi-2D) Mott-Hubbard 
antiferromagnetic insulators. The doping process initially produces 
the destruction of the antiferromagnetic ordering, giving place to 
a quantum spin-liquid disordered phase \cite{Anderson}. The two best 
studied examples are {\lco} (LSCO) and {\ybco} (YBCO) for which the 
N\'eel ordered ground state at $x=0$ is replaced by a quantum 
disordered state for $x_{c}\approx 0.02$  \cite{Borsa} and 
$x_{c}\approx 0.41$ \cite{Rossat-Mignod}, respectively, at
$T=0$.

It has long been recognized that the pure compounds are well 
described in terms of an $S=1/2$ quantum Heisenberg antiferromagnet 
(QHAF) on a square lattice. The long wavelength spin fluctuations 
of the latter, on the other hand, are described by the $O(3)$ 
quantum nonlinear sigma model (QNL$\sigma$M) in two space plus 
one time dimensions \cite{Haldane,CHN} whose Lagrangian density is
\beq
{\cal L}=
\frac{\rho_{s}}{2}
\left[\frac{1}{c^{2}}(\partial_{\tau}{\bf n})^{2} 
+(\nabla{\bf n})^{2}\right],
\eeq
where ${\bf n}$ is the order 
parameter field,
subject to the constraint ${\bf n}^{2}=1$,
and $\rho_{s}$ and $c$ are respectively the
spin-stiffness and spin-wave velocity.

The nonlinear sigma model possesses classical topologically 
nontrivial solutions called {\it skyrmions} whose energy is 
$E^{cl}_{s}=4\pi\rho_{s}$. For fully quantized skyrmions, on 
the other hand, there is a reduction of the skyrmion energy to
half of the above classical value \cite{Marino-PRB}
\beq
E_{s}=2\pi\rho_{s}.
\label{Es-Undoped}
\eeq
Having in mind that the skyrmion energy is reduced by 
quantum fluctuations, it is natural to expect it to be 
further reduced by the extra fluctuations introduced 
through doping. Furthermore, since in the framework of 
the NL$\sigma$M the skyrmion energy is proportional to 
the ground state magnetization we can use it as an order 
parameter for the quantum phase transition associated to 
the destruction of the antiferromagnetic state.

In this paper we revisit the continuum model proposed in 
\cite{Marino}, to describe the doping process in YBCO
at $T=0$, and extend it for including also the case 
of LSCO. The model predicts the creation of skyrmion 
topological defects precisely at the dopant's positions, 
as has been proposed earlier \cite{defects} (see also
\cite{Timm}). The formation
of such magnetic textures causes a reduction of the ground
state magnetization. As a consequence, the skyrmion energy 
itself is lowered and eventually vanishes at the N\'eel 
quantum critical point. By computing quantum defect correlation 
functions at $T=0$, we obtain analytical expressions for the 
skyrmion energy as a function of doping, $E_{s}(x)$, which 
allow us to plot the curves $M(x)\times x$, for both LSCO 
and YBCO, in good quantitative agreement with experiment. 
Subsequently, introducing finite temperature and interlayer 
coupling, we obtain the antiferromagnetic part of the phase 
diagram, namely the curve $T_{N}(x_{c}) \times x_{c}$. 
As we shall discuss below, our analysis is compatible with a 
picture in which the formation of stripes would occur in 
LSCO but not in YBCO.

{\it The doping process:} The chemical modification of 
parent compounds of the high-T$_c$ materials here considered produces the 
introduction of holes in the Oxygen orbitals in the 
layered {\cuoo} planes. In what follows, we shall 
determine how the skyrmion energy is modified from
(\ref{Es-Undoped}) due to the presence of such holes. 
We propose that the doped system can be described in 
terms of a bulk ($T=0$) $O(3)$ QNL$\sigma$M coupled to 
four-component fermion fields with dispersion relation 
determined by the shape of the Fermi surface. These represent 
the holes doped into the in-plane O$^{--}$ $p$-orbitals, 
whereas the nonlinear sigma field represents the spin 
density of the active electrons of the Cu$^{++}$ ions. 
For describing the coupling of the dopants to the 
Cu$^{++}$ spins, it will be convenient to make use of 
the CP$^{1}$ language, where ${\bf n}=z_{i}^{\dag}\vec{\sigma}_{ij}z_{j}$, 
with $\vec{\sigma}_{ij}$ being the Pauli matrices and 
$z_{i}, \ i=1,2$, complex scalar fields. Then, following 
\cite{Marino} (see also \cite{Wen-Shankar}),
we minimally couple the fermions to the
CP$^{1}$ fields. This is also 
consistent with previous results where a minimal coupling 
of fermions to CP$^{1}$ fields was obtained in the long 
wavelength regime of the spin-fermion model \cite{Kuebert}. 
Let us consider the cases of YBCO and LSCO separately.

{\bf a)} For YBCO we have an almost circular shape for the Fermi
surface, which is centered at ${\bf k} = 0$ \cite{Kampf},
see Fig. \ref{Tempdopfig1}a.
We can then use the dispersion
relation $\epsilon(k)=\sqrt{k^{2}v_{F}^{2}+(m^{*}v_{F}^{2})^{2}}$, 
with $m^{*}$ and $v_{F}$ being respectively the effective mass and 
Fermi velocity of the dopants. Observe that close to the Fermi level
of the doped system ($\epsilon_F > m^* v_F^2$) the above dispersion relation
behaves as $\epsilon (k) - \epsilon_F \simeq v_F (k -k_F)$ as expected.
This dispersion
relation corresponds to a Dirac kinetic term for the fermions and the
doped system can then be described by the partition function
\bea
{\cal Z}&=&\int{\cal D}[\bar{z},z,{\cal A}_{\mu},{\overline{\psi}},\psi]
{\;}\delta[\bar{z}z - 1]{\;}\delta[j^\mu - \Delta^\mu] \nonumber \\
&\times& \exp \left\{\int_{0}^{\infty}\rmd\tau\int\rmd^{2}{\bf x}
\left[2\rho_{s}
(D_{\mu}z_{i})^{\dag}(D^{\mu}z_{i})\right.\right.\nonumber\\
&+&\left.\left. \overline{\psi}(\rmi\partial\!\!\!\slash-\frac{m^{*}v_{F}}{\hbar}-
\gamma^{\mu}{\cal A}_{\mu})\psi+{\cal L}_{H}\right]\right\}
\label{CP1-Fermions}
\eea
where we allow for a residual Hopf term 
$
{\cal L}_{H}=(\theta/2)\varepsilon^{\mu\alpha\beta}
{\cal A}_{\mu}\partial_{\alpha}{\cal A}_{\beta},
$
which has been shown to exist in the presence of topological defects 
on the ground state \cite{Haldane2}. As usual,
$D_{\mu}=\partial_{\mu}+\rmi {\cal A}_{\mu}$, with 
$\partial_{\mu}=(\partial_{\tau}/c,\nabla)$, and ${\cal A}_{\mu}=
\rmi z_{i}^{\dag}\partial_{\mu}z_{i}$ is the Hubbard-Stratonovich 
CP$^{1}$ vector field.


%
\begin{figure}[h]
\centerline{\epsfxsize=8cm \epsffile{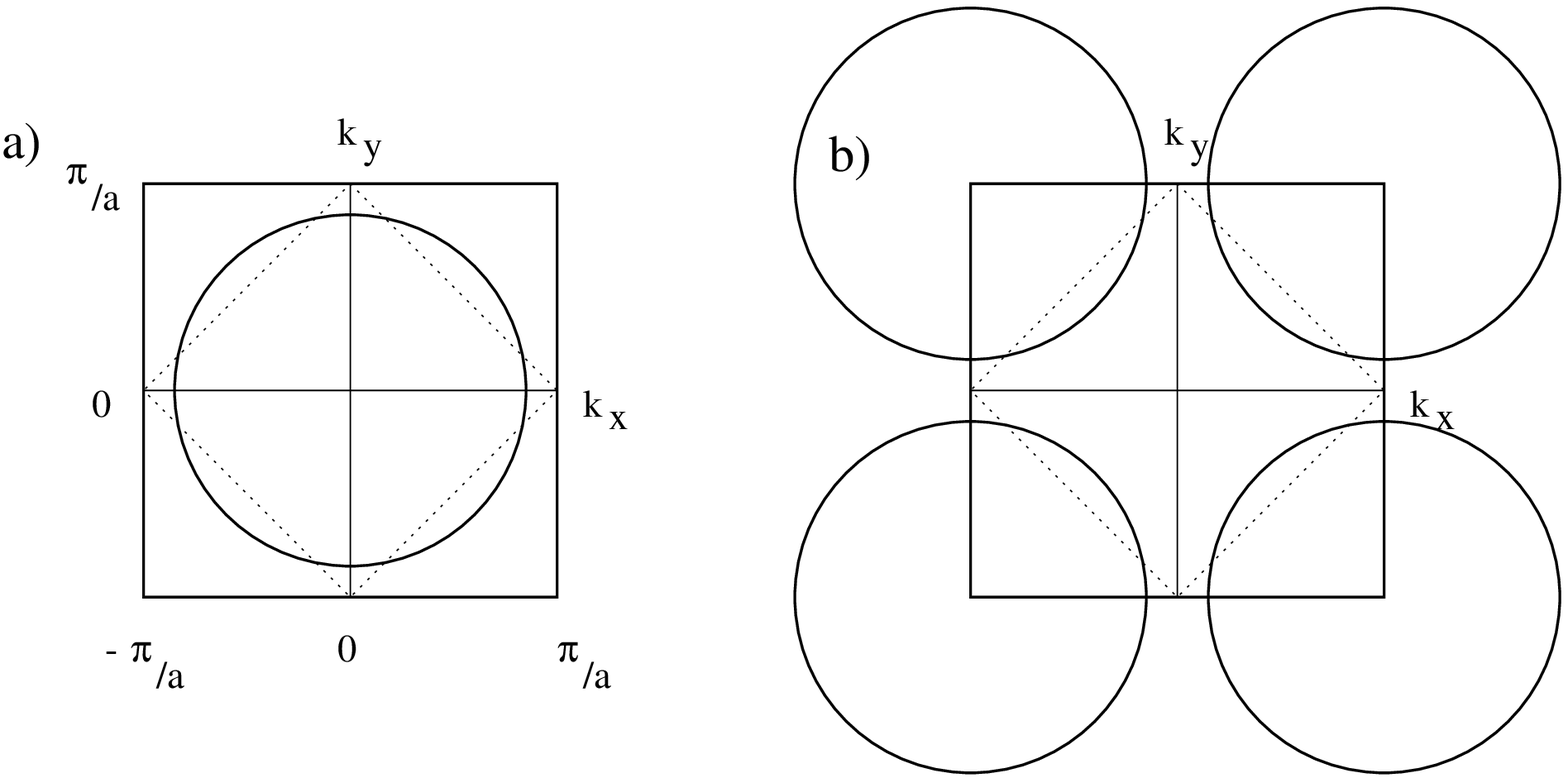}}
\vskip0.2cm
\caption{Approximate Fermi surfaces for:
a) {\ybco} and b) {\lco} \protect\cite{Kampf}.}
\label{Tempdopfig1}
\end{figure}
%

The second delta functional constraint in (\ref{CP1-Fermions}) 
is used in order to introduce the in-plane hole concentration 
parameter $\delta$. In its argument, $j^{\mu}=\bar{\psi}
\gamma^{\mu}\psi$ and $\Delta^{\mu}=4\delta\int_{X,L}^{\infty}
\rmd\xi^{\mu}\delta^{3}(z-\xi)$ for a dopant introduced at 
the position $X$ and moving along the line $L$. The factor 
of four in the definition of $\Delta^{\mu}$ accounts for the 
degeneracy of the adopted representation ($4$-component) for 
the Fermi fields. The zeroth component of the associated 
Lagrange multiplier will be the chemical potential. 

It has been shown in \cite{Marino}, that upon 
integration over the fields $\bar{z},z,{\overline{\psi}},\psi$, 
the resulting equation of motion for ${\cal A}_{0}$ is such that 
a skyrmion topological defect configuration concides with the 
dopant position at any time and $\pi\theta= 2\delta$. In other 
words, holes dress with skyrmions and we
see that indeed a Hopf term is required in (\ref{CP1-Fermions}) 
for nonzero doping. We stress that, because of the CP$^1$
constraint, a mass term is generated for the ${\cal A}_{\mu}$ field. 
Therefore, as a consequence of screening,
there will be neither statistical transmutation
nor the generation of spurious in-plane magnetic fields which 
would be inconsistent with muon spin relaxation 
experiments \cite{MSR}.

The parameter $\delta$ counts the number of holes in the {\cuoo} 
planes. This must be connected to the oxygen stoichiometry 
parameter $x$. For YBCO, it is known that the out of plane 
{\oco} chains play an important role in the process of doping. At 
low doping, $x\leq 0.18$, most of the holes go to the out of 
plane chains and the system can be considered as pure. 
In view of this we propose that the density of in-plane charge 
carriers is related to the oxygen stoichiometry by  $\delta=x-0.18$.

For evaluating the skyrmion energy $E_{s}$, we use the fact that 
skyrmions are topological defects whose quantum properties 
can be fully described by disorder field opertors $\mu$ 
\cite{NATO,Marino-PRB}. Using these, we are able to
calculate the large distance
behavior of the skyrmion correlation function, 
$\left<\mu^{\dag}(X)\mu(Y)\right> \rightarrow
e^{-(E_{s}/\hbar c)|X-Y|}/|X-Y|^{\nu}$, from which we can 
extract the skyrmion energy $E_{s}$ \cite{Marino-PRB}. For 
the pure system we have that $E_{s}$ is given by (\ref{Es-Undoped}).
For the partition function (\ref{CP1-Fermions}),
we obtain, after integration over $\bar{z},z,$ and $\overline{\psi},\psi$ 
fields, \cite{Marino}
\beq
E_{s}(\delta)= 2\pi\rho_{s} \left (1-
\frac{\gamma\hbar c}{\pi a_{D}\rho_{s}}\delta^{2}\right ),
\label{Es-YBCO}
\eeq
where $\gamma=\frac{32\pi(9\pi^{2}-16)}{(\pi^{2}+16)^{2}}=10.9398$ is a 
numerical factor that comes from the integration over the fermions, and
$a_{D}\equiv a/\sqrt{2}=2.68$ {\AA}, the minimal distance between two oxygen atoms,
is the lattice spacing for dopants. The above skyrmion energy can be put 
in the form $E_{s}=2\pi\rho_{s}(\delta)$ if we define an effective $\delta$ 
dependent spin-stiffness 
\beq
\rho_{s}(\delta)=\rho_{s}\left(1-\frac{\gamma\hbar c}
{\pi a_{D}\rho_{s}}\delta^{2}\right).
\label{Rho-of-Delta-YBCO}
\eeq
We then conclude, after comparing (\ref{Es-YBCO}) with
(\ref{Es-Undoped}), that the doped system can be described by a 
QNL$\sigma$M with a generalized $\delta$ dependent spin-siffness 
given by (\ref{Rho-of-Delta-YBCO}).

We can immediately obtain the
reduced sublattice magnetization as a function of doping as:
$M(x)/M(0)=\sqrt{\rho_{s}(x)/\rho_{s}}$.
This is plotted in Fig. \ref{Tempdopfig2}
for $\hbar c=1.00 \pm 0.05$ eV {\AA} and $\rho_{s}=0.069$ eV \cite{Kampf}. 
We see that our theoretical prediction is in good agreement with
experiment and $\rho_{s}(\delta)$ vanishes for
\beq
\delta_{c}=\sqrt{\frac{\pi\rho_{s}a_{D}}{\gamma\hbar c}}.
\eeq
Using the above experimental input,
we obtain $\delta_{c}=0.23\pm0.03$ or $x_{c}=0.41\pm0.03$ 
for the quantum critical point.
 

%
\begin{figure}[h]
\centerline{\epsfxsize=6cm \epsffile{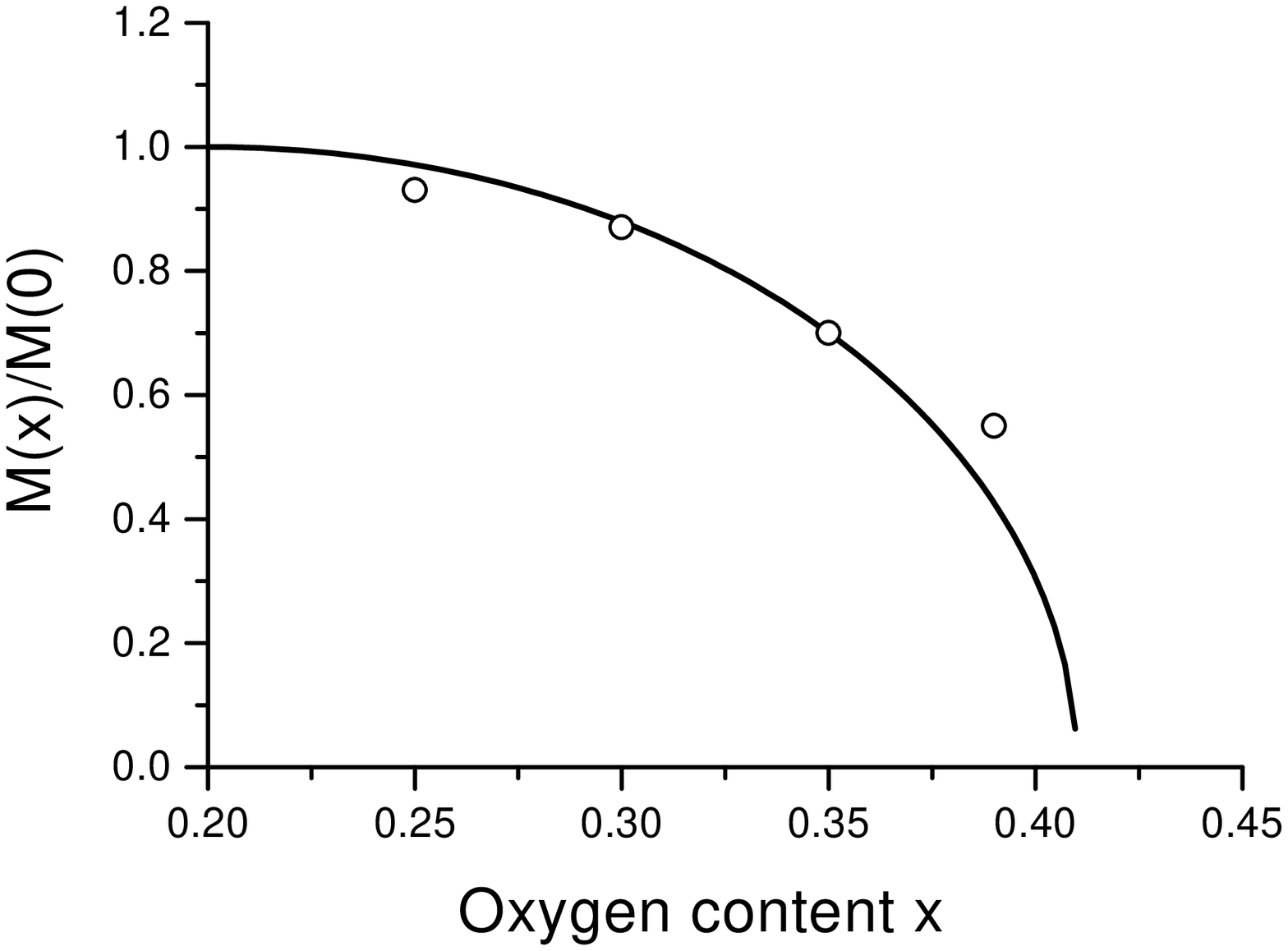}}
\caption{Reduced sublattice magnetization for 
{\ybco}. Experimental data from \protect\cite{Bucci}.}
\label{Tempdopfig2}
\end{figure}
%

{\bf b)} For the case of LSCO, the Fermi surface is given 
approximately by the one of Fig. \ref{Tempdopfig1}b, which has 
also been observed in 
ARPES \cite{ARPES}. We shall use the fact that the Fermi surface of 
LSCO can be obtained by shifting the one of YBCO towards 
$Q=(\pi/a,\pi/a)$ and symmetry related points in the Brillouin 
zone. This leads to the new dispersion relation 
$\epsilon(k)=\sqrt{[(k_{x}\pm\pi/a)^{2}+(k_{y}\pm\pi/a)^{2}]v_{F}^{2}+
(m^{*}v_{F}^{2})^{2}}$, which corresponds to a kinetic term for the
fermions in (\ref{CP1-Fermions}) modified by a shifted derivative term.
Even though the partition function remains unchanged, this shift will have
a nontrivial effect on the skyrmion correlation function and energy because
of the constraint on the fermionic density of states. The computation of 
the skyrmion correlation function proceeds as before
\cite{Marino,Marino-PRB} and we now obtain the
following expression for the skyrmion energy
\beq
E_{s}(\delta)= 2\pi\rho_{s} \left (1-
\frac{\gamma\hbar c}{\pi a_{D}\rho_{s}}(4\delta)^{2}
-\frac{2\sqrt{2}\hbar c}{a\rho_{s}}(4\delta) \right ).
\label{Rho-of-Delta-LCO}
\eeq

There are two important differences between (\ref{Rho-of-Delta-LCO}) 
and (\ref{Es-YBCO}). The first one is a linear $\delta$ dependence
in (\ref{Rho-of-Delta-LCO}) which is absent in (\ref{Es-YBCO}).
This was generated by the shift in the dispersion relation and is a 
consequence of the fact that the Fermi surface of LSCO is not 
centered at ${\bf k}=0$ in the Brillouin zone. The second one is an 
extra factor of four multiplying $\delta$ in (\ref{Rho-of-Delta-LCO}), 
which is needed to account for the four branches of the Fermi surface 
of this compound.

The above skyrmion energy suggests, in accordance to what has been done
for the case of YBCO, that the doped system could presumably be described 
by a QNL$\sigma$M with stiffness 
\beq
\rho_{s}(\delta)=\rho_{s}\left(1-\frac{\gamma\hbar c}{\pi a_{D}\rho_{s}}
(4\delta)^{2}-\frac{2\sqrt{2}\hbar c}{a\rho_{s}}(4\delta)\right).
\label{Stiffness-No-Stripes}
\eeq
However, the magnetization that would follow from this, according to
the same procedure adopted for the case of YBCO,
has a faster decrease with doping than the one observed experimentally,
as we can infer from the solid line in from Fig. \ref{Tempdopfig3}.
This can be interpreted as a sign of stripes formation in LSCO 
\cite{Carretta}. Following \cite{Castro-Neto}, we observe that
the presence of stripes produces an anisotropy which allows us to
write $\rho_{s}^{\prime}(\delta)=\sqrt{\rho_{s}^{x}(\delta)\rho_{s}^{y}(\delta)}$, 
where $\rho_{s}^{x}(\delta)\neq\rho_{s}^{y}(\delta)$. Assuming that in LSCO
stripes are parallel to the $y$-axis, as the experimental results in
\cite{Cheong} suggest, we propose that only the $x$-component of
the spin-stiffness would be affected by doping, implying  that  
$\rho_{s}^{y}=\rho_{s}$ and that $\rho_{s}^{x}(\delta)$ is given 
by (\ref{Stiffness-No-Stripes}). The resulting NL$\sigma$M has an 
effective spin-stiffness 
\beq
\rho^{\prime}_{s}(\delta)=\rho_{s}
\sqrt{1-\frac{\gamma\hbar c}{\pi a_{D}\rho_{s}}
(4\delta)^{2}-\frac{2\sqrt{2}\hbar c}{a\rho_{s}}(4\delta)}.
\label{Stiffness-Stripes}
\eeq
Now the corresponding sublattice
magnetization does agree with the experimental data as can be seen from the
dashed line in Fig. \ref{Tempdopfig3}.
Both plots in this figure are obtained by using the experimental input
$\hbar c=0.75 \pm 0.03$ eV {\AA} and $\rho_{s}=0.051$ eV \cite{Kampf}.
We have also used the well known fact that for LSCO, $x = \delta$.
Interestingly, the location of the quantum
critical point is not affected by the formation of stripes. 
In fact, both $\rho_{s}(\delta)$ and $\rho^{\prime}_{s}(\delta)$
vanish at
\beq
\delta_{c}=\frac{1}{4}
\frac{\sqrt{(2\sqrt{2}\hbar c/a)^{2}+4 \gamma\hbar c\rho_{s}/\pi a_{D}}-
(2\sqrt{2}\hbar c/a)}{2 (\gamma\hbar c/\pi a_{D})}.
\label{Critical-Curve}
\eeq
which, for the above experimental input, yields the value
$x_{c}=\delta_{c}=0.020\pm0.003$ for the quantum 
critical point. We also remark that using the effective 
spin-stiffness (\ref{Stiffness-Stripes}), implied by the
stripes picture, the reduced sublattice magnetization
$M(x)/M(0)=\sqrt{\rho^{\prime}_{s}(x)/\rho_{s}}$
can be written in the form $M(x)/M(0)=(1-x/x_{c})^{1/4}$, 
thereby producing a critical exponent of $0.25$, which is 
very close to the value of $0.236$, empirically obtained by 
Borsa {\it et al.} \cite{Borsa}.


%
\begin{figure}[h]
\centerline{\epsfxsize=6cm \epsffile{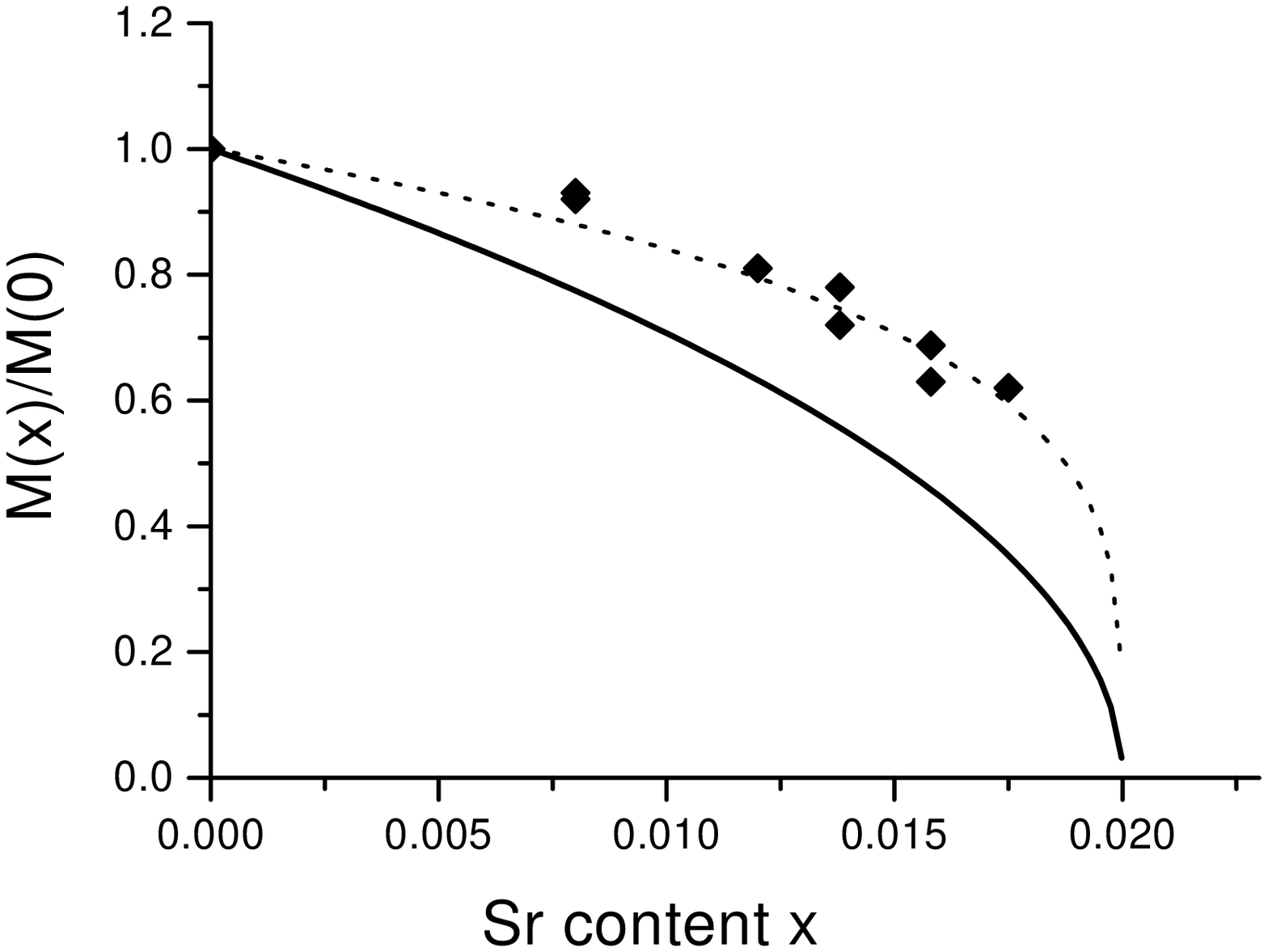}}
\caption{Reduced sublattice magnetization for {\lco}.
Experimental data from \protect\cite{Borsa}.}
\label{Tempdopfig3}
\end{figure}
%

{\it The phase diagram:} In order to obtain the critical line for the
destruction of AF order in the
$T\times x$ phase diagram we must consider a nonzero interlayer 
coupling $J_{\perp}$. For this purpose, we shall use the partition
function \cite{Katanin}
\bea
{\cal Z} & = & \int{\cal D} {\bf n}_{i} 
\exp \left\{-\frac{\rho_{s}}{2\hbar}
\int_{0}^{\hbar\beta}\rmd\tau\int\rmd^{2}{\bf x}
\sum_i\left[\frac{1}{c^{2}}
(\partial_{\tau} {\bf n}_{i})^{2} \right. \right. \nonumber \\
& + & \left. \left. (\nabla {\bf n}_{i})^{2}+
\alpha({\bf n}_{i+1}-{\bf n}_{i})^{2} 
\right] \right\} {\;}\delta[{\bf n}_{i}^{2}-1],
\label{Part-Func}
\eea
where $\alpha=(1/a^{2})J_{\perp}/J_{||}$, with $J_{\perp}$ and $J_{||}$
being respectively the interlayer and intralayer couplings of the 
underlying microscopic quasi-2D QHAF and $\beta = 1/T$ ($k_B=1$).
Contrary to the strictly 2D case now a finite value for the N\'eel
temperature is obtained to order $1/N$ in a large $N$ expansion
\cite{Katanin}:
\beq
T_{N}=4\pi\rho_{s}
\left[\ln{\left(\frac{2T_{N}^{2}}{\alpha(\hbar c)^{2}}\right)}
+3\ln{\left(\frac{4\pi\rho_{s}}{T_{N}}\right)}-0.0660
\right]^{-1}.
\label{Neel-Temperature}
\eeq
Now, if we replace 
in (\ref{Neel-Temperature}) the spin-stiffness $\rho_{s}$ by our $\delta$ 
dependent expressions (\ref{Rho-of-Delta-YBCO}) and
(\ref{Stiffness-Stripes}), respectively for YBCO and LSCO,
and using $J_{\perp}/J_{||}\simeq 5\times 10^{-5}$, 
we obtain the phase diagrams plotted in Figs. \ref{Tempdopfig4}
and \ref{Tempdopfig5}.


%
\begin{figure}[h]
\centerline{\epsfxsize=6cm \epsffile{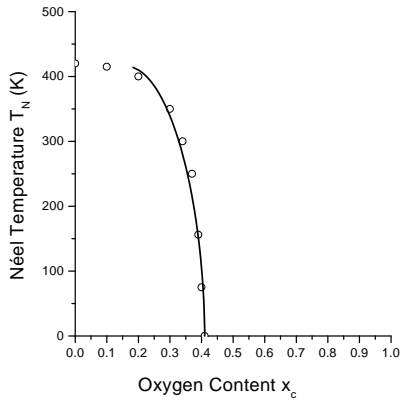}}
\caption{Antiferromagnetic part of the phase diagram for {\ybco}.
For $x_c < 0.18$, we assume a fixed $T_N = 420 K$.
Experimental data from \protect\cite{Rossat-Mignod}.}
\label{Tempdopfig4}
\end{figure}
%

%
\begin{figure}[h]
\centerline{\epsfxsize=6cm \epsffile{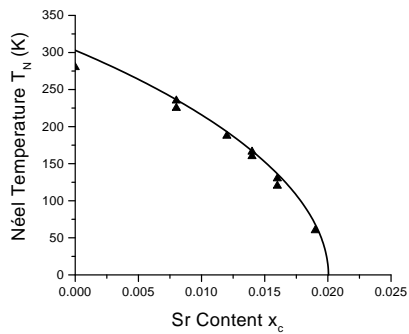}}
\caption{Antiferromagnetic part of the phase diagram for {\lco}. 
Experimental data from \protect\cite{Borsa}.}
\label{Tempdopfig5}
\end{figure}
%

In summary, our results indicate that the antiferromagnetic 
phase of the high-T$_c$ cuprates can be correctly described in terms 
of a generalized NL$\sigma$M with an effective, doping dependent, 
spin-stiffness which carries all the information about the quantum 
fluctuations introduced by the dopants. This can be inferred from the
quantum skyrmion energy which is evaluated after
integration over the dopants in a model without adjustable parameters.
Also, the results point towards
a picture in which charged stripes presumably occur in LSCO, leading 
to a slower decrease of the sublattice magnetization as a function 
of doping in comparison to the case where they are absent, see Fig. 
\ref{Tempdopfig3}. Apparently, the only effect of such phenomenon is 
to divide by two the quantum critical exponent of the sublattice
magnetization (or the spin-stiffness)
without affecting the location of the quantum critical point. Conversely, 
our results indicate that such phenomenon is absent in YBCO as our 
model (\ref{CP1-Fermions}) was able to correctly reproduce the 
zero temperature data for $M(x)/M(0)$, see Fig. \ref{Tempdopfig2}. 
This is consistent with density matrix renormalization group calculations 
in the framework of the $n$-leg ladder $t-t^{\prime}-J$ model 
\cite{White-Scalapino}.

As a final comment we would like to remark that for the case of
Bi$_2$Sr$_2$CaCu$_2$O$_{8+x}$ we expect that our result 
(\ref{Stiffness-No-Stripes}) shall correctly reproduce the 
experimental data without the scaling modifications 
done for LSCO. Indeed this material, while having a Fermi surface 
similar to LSCO \cite{Ding}, does not seem to present the formation 
of stripes.


We are indebted to A. H. Castro Neto, A. A. Katanin, B. Koiller, C. K\"ubert, 
E. Miranda and J. Schmalian for many useful comments. We also acknowledge P. 
Carretta for pointing out Ref. [16]. E.C.M. was partially supported by CNPq 
and FAPERJ. M.B.S.N was supported by FAPERJ.

\end{narrowtext}


\end{document}